\documentclass[aip,amssymb, amsmath]{revtex4-1}
\usepackage{graphicx, float}
\usepackage{subcaption}
\usepackage{adjustbox}
\usepackage[utf8]{inputenc}
\usepackage[T1]{fontenc}
\usepackage{etoolbox}
\usepackage{dcolumn}
\usepackage{bm}
\usepackage{color}
\draft

\begin{document}

\title{The Impact of Collisionality on the Runaway Electron Avalanche during a Tokamak Disruption}

\author{J. Arnaud} 
\affiliation{University of Florida}
\author{C.J. McDevitt}
\affiliation{University of Florida}

\date{\today}

\begin{abstract}
The exponential growth (avalanching) of runaway electrons (REs) during a tokamak disruption continues to be a large uncertainty in RE modeling. The present work investigates the impact of tokamak geometry on the efficiency of the avalanche mechanism across a broad range of disruption scenarios. It is found that the parameter $\nu_{*, crit}$ describing the collisionality at the critical energy to run away delineates how toroidal geometry impacts RE formation. In particular, utilizing a reduced but self-consistent description of plasma power balance, it is shown that for a high-density deuterium-dominated plasma, $\nu_{*, crit}$ is robustly less than one, resulting in a substantial decrease in the efficiency of the RE avalanche compared to predictions from slab geometry. In contrast, for plasmas containing a substantial quantity of neon or argon, $\nu_{*, crit} \gtrsim 1$, no reduction of the avalanche is observed due to toroidal geometry. This sharp contrast in the impact of low-versus high-Z material results primarily from the relatively strong radiative cooling from high-Z impurities, enabling the plasma to be radiatively pinned at low temperatures and thus large electric fields, even for modest quantities of high-Z material.

\end{abstract}

\maketitle 

\section{Introduction}
\label{Introduction}
Disruptions pose a significant threat to the success of the tokamak concept in magnetic fusion. During a tokamak disruption, nearly all of the plasma's thermal energy is lost, leading to a drastic increase in the plasma resistivity and a subsequent spiking of the electric field. The force exerted on an electron by the resulting electric field can greatly exceed the electron's collisional drag, thus enabling the acceleration of electrons to relativistic energies \cite{Connor:1975}. The number of these so-called runaway electrons (REs) can in turn be amplified by a process referred to as `avalanching' \cite{Sokolov:1979}. Here, a pre-existing RE undergoes a large-angle collision with a low-energy electron. If the energy transferred during this collision is sufficient to scatter the initially cold electron to high enough energy to run away (often referred to as a secondary electron), this will lead to the growth of the initial RE population. This process has been observed in existing tokamaks to enable the transfer of plasma current to the RE population, with the potential for mega-Ampere RE currents being inadvertently generated in reactor-scale tokamak plasmas \cite{Rosenbluth:1997}. Moreover, the energy of RE beams is typically in the MeV range, which implies the possibility of subsurface damage to plasma-facing components, further raising the importance of RE mitigation.

Understanding the efficiency of the avalanche mechanism is important for predicting the amount of RE current that can be expected to form during a tokamak disruption. While the impact of partially ionized impurities on avalanche generation has been extensively studied \cite{Martin:2015, mcdevitt2019avalanche, hesslow2018generalized}, substantial uncertainties remain regarding the impact of tokamak geometry on the avalanche mechanism. In particular, the magnetic field strength in a tokamak device scales as $B \propto 1/R$, where $R$ is the major radius. Electrons with modest values of pitch $\xi \equiv p_\Vert / p$, are susceptible to magnetic trapping, leading to the formation of banana orbits. In the presence of a parallel electric field, electrons undergoing banana orbits will be accelerated on one leg of the orbit and decelerated on the returning leg, leading to a negligible acceleration of the electron after a complete banana orbit. Since a substantial fraction of secondary electrons are born with small values of the pitch $\xi \equiv p_\Vert / p$, the presence of a trapped region can substantially reduce the efficiency of the avalanche mechanism \cite{Rosenbluth:1997,Chiu:1998, Harvey:2000, hoppe2021dream}. 

A caveat to the above picture, however, is that it requires the electron to have sufficiently low collisionality such that the electron is able to complete a full bounce orbit before being collisionally de-trapped. While MeV electrons are expected to be in the collisionless regime for parameters characteristic of a tokamak disruption, the collisionality at the critical energy for an electron to run away is expected to be far larger. In particular, defining the critical energy for an electron to run away by the energy at which the drag and electric field terms balance, this energy can be approximated by $m_e v^2_{crit}/2 \sim \left( m_e c^2 / 2\right) \left( E_c /E_\Vert\right)$, where we have assumed $E_\Vert/E_c \gg 1$, $E_c \equiv m_ec/(e\tau_c)$ is the Connor-Hastie electric field \cite{Connor:1975}, and $\tau_c \equiv 4\pi \epsilon^2_0 m^2_e c^3/ \left( e^4 n_e \ln \Lambda \right)$ is the relativistic collision time. During a tokamak disruption, this energy can obtain values ranging from several to tens of keV, depending on the specific plasma composition present during the disruption. Noting that the collisionality scales inversely with energy, electrons near $v_{crit}$ often will not be in the banana collisionality regime, potentially negating electron trapping effects on the avalanche growth rate \cite{mcdevitt2019runaway}.

The present paper aims to employ an idealized but self-consistent model of a tokamak disruption to identify the collisionality regime characterizing runaway electron formation across a broad range of disruption scenarios, with particular emphasis on the avalanche mechanism of RE formation. Specifically, defining the dimensionless collisionality parameter by $\nu_* \equiv \tau_{bounce}/\tau_{dt}$ , where $\tau_{bounce} = qR_0/(\epsilon^{1/2}v)$ and $\tau_{dt} = \varepsilon/\nu_D(v)$ are the bounce and collisional de-trapping times, respectively \cite{Helander-Sigmar:book}. The collisionality $\nu_*$ can be further expressed as
\begin{equation}
\nu_* = \left(\frac{qR_0}{a\varepsilon^{3/2}}\right)\left(\frac{c}{v}\right)\left(\frac{a}{c\tau_c}\right)(\tau_c\nu_D(v))
\label{nu_*}
\end{equation}
where $a$ is the minor radius, $R_0$ is the major radius, $\varepsilon = r/R_0$ is the inverse aspect ratio, $q$ is the safety factor, and $\nu_D (v)$ is the pitch-angle scattering rate. Noting that $c\tau_c/a \propto 1/(n_e\ln\Lambda)$, where $n_e$ is the free electron density and $\ln\Lambda$ is the Coulomb logarithm, the collisionality at the critical speed $v_{crit}$ for an electron to run away scales as
\begin{equation}
\nu_{*,crit} \propto n_e\ln\Lambda\left(\frac{c}{v_{crit}}\right)(\tau_c\nu_D(v_{crit}))
\label{nuscale}
\end{equation} 
where we have omitted the geometric parameter $qR_0 /\left( a\varepsilon^{3/2} \right)$ for simplicity. Here, $\nu_{*,crit}$ depends sensitively on the plasma density, pitch-angle scattering frequency, and $v_{crit}$. During a tokamak disruption, these three quantities can span a broad range of values, depending on the disruption mitigation scheme pursued and the device under consideration. For example, the large inductive electric field present during a disruption is the result of a sharp decrease in the plasma temperature $T_e$ driving an abrupt rise in the plasma resistivity. The decrease in $T_e$ is dependent on the plasma composition ($n_{impurity}, n_D$), which directly relates $E_\Vert$ to the quantity and species of atoms injected into the plasma as part of a disruption mitigation scheme or that is released from the wall due to plasma-material interaction. The primary motivation of this paper will thus be to quantify how distinct plasma compositions impact the value of $\nu_{*,crit}$ and thus the efficiency of the RE avalanche.

The remainder of this paper is organized as follows: Section \ref{self-consistent} describes the self-consistent formulation for evaluating $\nu_{*,crit}$ and evaluates it across a broad range of plasma compositions. The relativistic drift kinetic solver used to evaluate the RE growth rate is briefly introduced in Sec. \ref{RAMc}. Section \ref{results} evaluates the avalanche growth rate for a range of disruption scenarios. Conclusions are given in Sec. \ref{discuss}.

\section{Self-Consistent Evaluation of the Collisionality at the Critical Energy}
\label{self-consistent}

\begin{figure}
\centering
\begin{subfigure}{.45\textwidth}
\caption{}
\includegraphics[width=\linewidth]{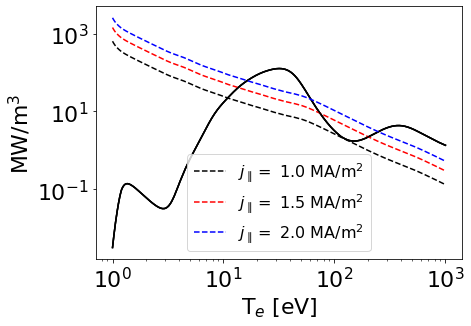}
\end{subfigure}%
\begin{subfigure}{.45\textwidth}
\caption{}
\includegraphics[width=\linewidth]{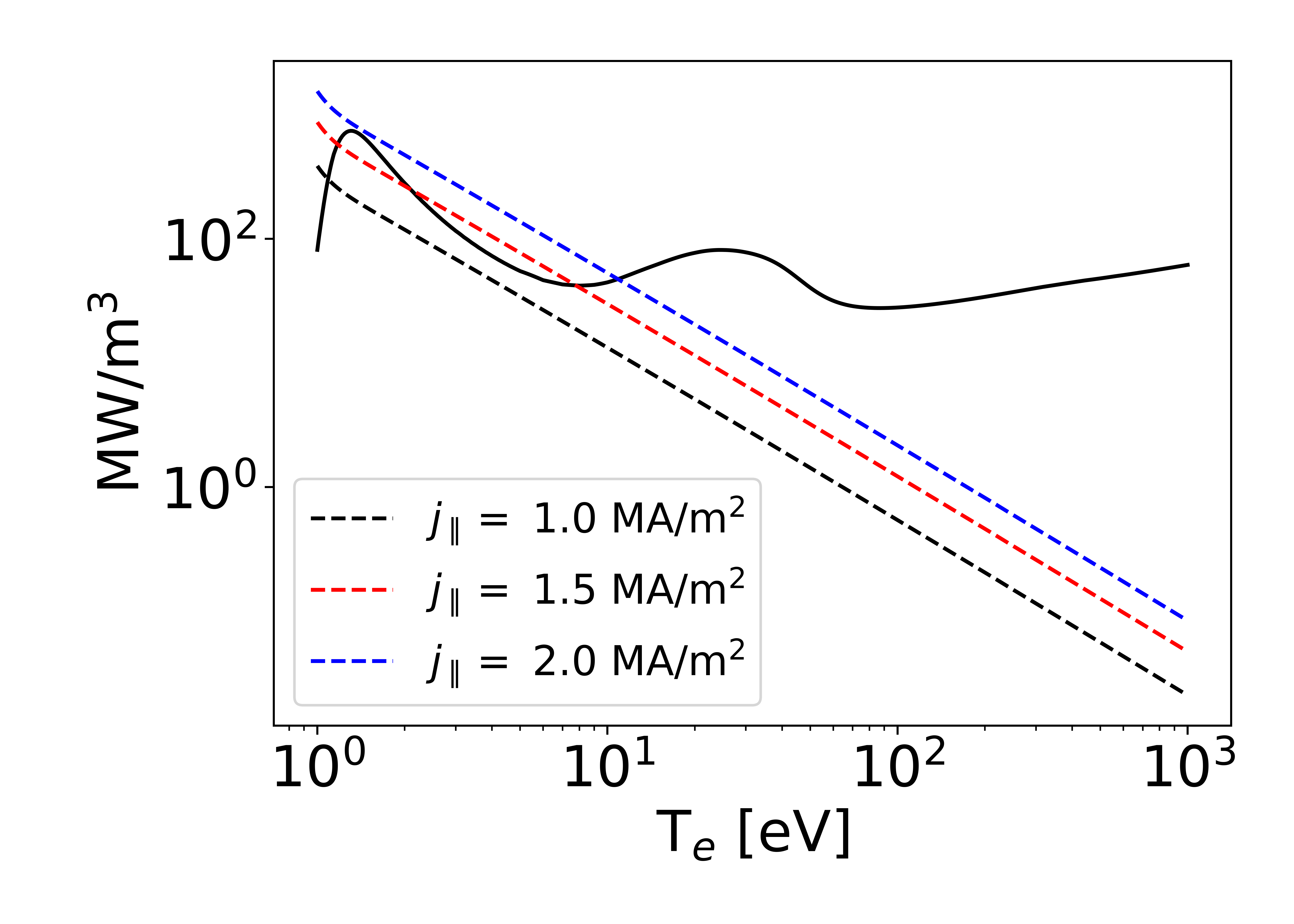}
\end{subfigure}
\begin{subfigure}{.45\textwidth}
\caption{}
\includegraphics[width=\linewidth]{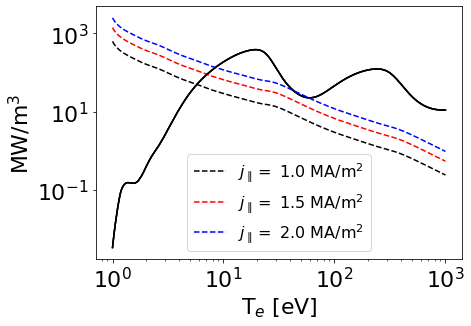}
\end{subfigure}%
\begin{subfigure}{.45\textwidth}
\caption{}
\includegraphics[width=\linewidth]{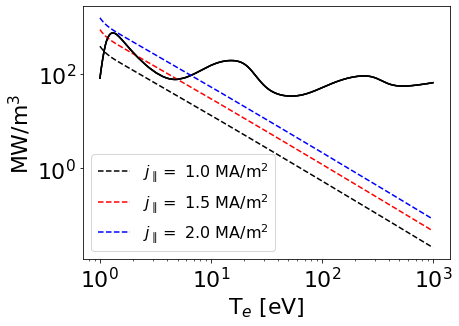}\end{subfigure}
\caption{0-D Power balance model for a plasma containing neon [panels (a) and (b)] and argon [panels (c) and (d)] impurities, where the solid lines are the radiative cooling ($S_{rad}$), and the dashed lines are the ohmic heating ($\eta j_{\Omega}^2$). The densities used were $n_D$ = 1 $\times$ 10$^{13}$ cm$^{-3}$ and n$_{impurity^{1+}}$/n$_D$ = 0.1 for (a) and (c), whereas panels (b) and (d) used $n_D$ = 1 $\times$ 10$^{16}$ cm$^{-3}$ and n$_{impurity^{1+}}$/n$_D$ = 10$^{-5}$.}
\label{MultRoots}
\end{figure}

\subsection{Collisional Radiative Model to Evaluate $T_e$ and $\overline{Z_j}$}
\label{coll-rad} 
During the thermal quench (TQ) phase of a tokamak disruption, where large regions of magnetic field stochasticity are expected, the plasma primarily cools by heat conduction along the open magnetic field lines to a temperature of roughly a hundred eV\cite{ward1992impurity}, after which radiation further cools the plasma. Losses from this latter mechanism will eventually be balanced by the Ohmic heating of the resistive plasma, resulting in a saturated post-TQ plasma temperature $T_e$. In order to self consistently evaluate $T_e$ for a given plasma composition, a power balance model must be employed. Here, data from the collisional-radiative models FLYCHK \cite{chung2005flychk} and ATOMIC \cite{fontes2015alamos} are used, where we consider a steady state 0-D power balance between the Ohmic heating of the bulk plasma ($j_{\Omega}^2$) and the radiative cooling ($S_{rad}$). In addition to radiative losses, these atomic data sets are used to evaluate the average charge state of the impurity $\overline{Z}_{impurity}$ (neon or argon) and deuterium $\overline{Z}_{D}$ ions, and $n_e$ is evaluated through quasineutrality. Two representative cases are shown in Fig. \ref{MultRoots} for singly charged neon [panels (a) and (b)] and argon [panels (c) and (d)] impurities, where Figs. \ref{MultRoots}(a) and (c) represent a scenario where a significant amount of impurities are present  ($n_{impurity}/n_D \approx 0.1$) and Figs. \ref{MultRoots}(b) and (d) represent a deuterium dominant plasma ($n_D/n_{impurity} \gg 1$). Depending on the current density and plasma composition, stable roots may exist at $\sim$ 1 eV, $\sim$ 5–30 eV, and $\sim$ 100–300 eV; however, we are only interested in roots with temperatures below the regime where heat conduction along stochastic magnetic field lines is subdominant. Hence, we will not consider the high temperatures roots with temperatures in the range $100-300\;\text{eV}$. Moreover, for plasmas with large quantities of deuterium [$n_D \sim 10^{16}\;\text{cm}^{-3}$, Figs. \ref{MultRoots}(b) and (d)], the $\sim$ 1 eV root enables the deuterium species to begin to recombine, thus defining a critical deuterium density above which the plasma is expected to be in a low ionization state. For these high deuterium density cases, where multiple roots between $\sim$ 1 eV and $\sim$ 10 eV exist, the larger of the two roots is used since we are considering scenarios where the plasma is cooling from higher temperatures. If the deuterium density is further increased or if the current density is reduced, the higher temperature root will be removed, leaving only the root at one to two eV.

Identifying the precise criteria for accessing this low-temperature root requires the incorporation of physics beyond the idealized power balance analysis discussed above. In particular, the power balance curves shown in Fig. \ref{MultRoots} were evaluated under the assumption of an optically thin plasma. While this is expected to be a good approximation for the majority of plasma compositions under consideration, recent work has noted that for sufficiently high deuterium densities, the plasma will become opaque to Lyman radiation, leading to a substantial reduction of radiative losses $S_{rad}$ for deuterium \cite{vallhagen2022effect}. While this reduction of radiative losses postpones the density at which deuterium recombines, this effect only impacts plasmas with exceptionally high deuterium densities, which make up a small portion of the cases treated in this analysis. We will thus employ the assumption of an optically thin plasma throughout this paper, where finite opacity effects will be left for future work.

\subsection{Electric Field Evaluation}

\begin{figure}
\centering
\begin{subfigure}{.45\textwidth}
\caption{}
\includegraphics[width=\linewidth]{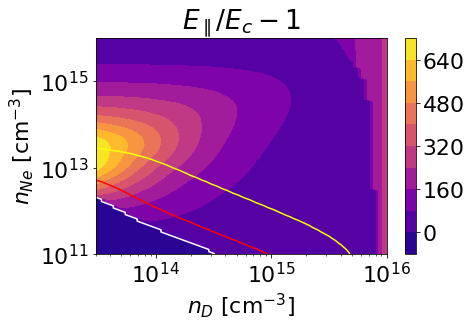}
\end{subfigure}%
\begin{subfigure}{.45\textwidth}
\caption{}
\includegraphics[width=\linewidth]{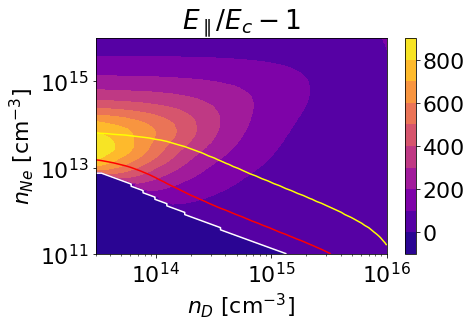}
\end{subfigure}
\begin{subfigure}{.45\textwidth}
\caption{}
\includegraphics[width=\linewidth]{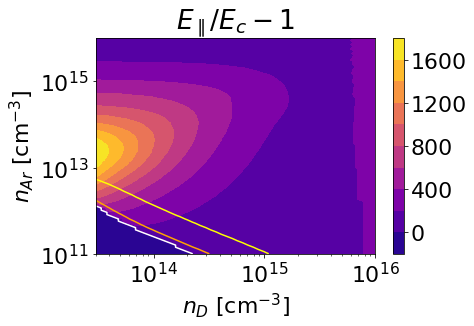}
\end{subfigure}%
\begin{subfigure}{.45\textwidth}
\caption{}
\includegraphics[width=\linewidth]{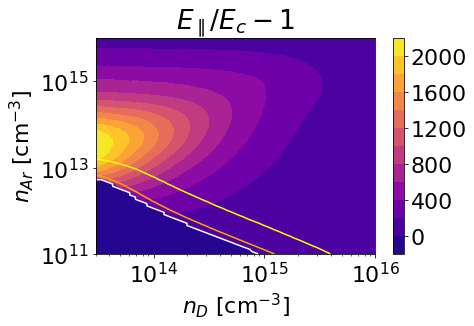}
\end{subfigure}
\caption{Post-TQ electric fields, normalized to $E_c$ for neon [panels (a) and (b)] and argon [panels (c) and (d)], where (a) and (c) are for $j_{\parallel}$ = 1 MA/m$^2$ and (b) and (d) are for  $j_{\parallel}$ = 2 MA/m$^2$. Other parameters used were an inverse aspect ratio of $\epsilon \equiv a /R_0 = 1/3$, a safety factor of $q = 2.1 + 2(r/a)^2$, a minor radius of  $a$ = 200 cm, and a magnetic field of $B$ = 5.3 T. The yellow, orange, and red contour lines represent $T_e$ = 10, 17, and 20 eV, respectively. The white contour represents electric fields below the Connor Hastie threshold electric field.}
\label{EoverEc}
\end{figure}

Once $T_e$, $n_e$, and $\overline{Z}_{impurity}$ are evaluated across a range of plasma compositions, $E_\Vert$ is obtained from Ohm's law: $E_\Vert = \eta_{s} j_\Vert$, where $\eta_{s}$ is the Spitzer resistivity \cite{cohen1950electrical} and $j_\Vert$ is the local current density. Here we are interested in scenarios during the early phase of a disruption where the RE current $j_{RE}$ is negligible, hence we neglect its impact on Ohm's law. Figure \ref{EoverEc} shows the electric fields evaluated from the power balance for argon and neon impurities and two different values of $j_\Vert$.  The yellow, orange, and red contours are for $T_e$ = 10, 17, and 20 eV, respectively. While the current decay time in ITER will depend on a range of factors~\cite{Hender:2007}, plasmas with temperatures of approximately ten to twenty eV will have current decay times near the ITER target range of 50–150 ms~\cite{lehnen2015disruptions}, and will thus be the primary focus of this analysis. We can understand the behavior of $E_\Vert/E_c$ by noting that $E_\Vert/E_c \propto \eta_s j_\Vert/n_e \propto Z_{eff}T_e^{-3/2}j_\Vert/n_e$ and considering the case of neon at a given $j_\Vert$ and $T_e$ [Fig. \ref{EoverEc} (b)].  As $n_D$ decreases along a given temperature contour (say, the yellow 10 eV contour), $E_\Vert/E_c$ increases significantly, which is due to the drop in the free electron density $n_e$ and the increase in $Z_{eff}$. Another feature present is a second large $E_\Vert/E_c$ region at deuterium densities of $n_D \approx 10^{16}$ cm$^{-3}$ in Figs. \ref{EoverEc}(a) and (c). The corresponding temperature at these deuterium densities is roughly an eV, thus driving recombination and lowering $n_e$ which decreases $E_c$. Once the current density is increased from 1 MA/m$^2$ to 2 MA/m$^2$, this second large $E_\Vert/E_c$ region at large deuterium densities is no longer present since the critical deuterium density at which the plasma recombines is increased. The impact of the choice of impurities on $E_\Vert/E_c$ is also shown in Figure \ref{EoverEc}. Comparing Figures \ref{EoverEc} (a) and (c), it can be seen that for argon, the $T_e$ = 10–20 eV channel is generally at lower deuterium and argon densities than neon, and the width is narrower for argon. As a result, argon impurities produce larger $E_\Vert/E_c$ in the region where 10 eV < $T_e$ < 20 eV, since  $E_c \propto n_e$. Furthermore, argon impurities globally increase $E_\Vert/E_c$ across all deuterium and argon densities, resulting from more efficient radiative cooling than neon. Specifically, comparing Figures \ref{MultRoots} (a) and (c), the radiative cooling for argon is stronger than that for neon, where at a given deuterium and argon density, the temperature root is lower for argon, thus increasing $E_\Vert$ in comparison to neon.

\subsection{\label{sec:CRE}Collisionality at the RE Critical Energy}
The specific form used to evaluate the collisionality at the critical energy $v_{crit}$ is 
\begin{equation}
\nu_{*, crit} = \left(\frac{qR_0}{\epsilon^{3/2}}\right)\left(\frac{\nu_D \left( v_{crit} \right)}{v_{crit}}\right)
, \label{eq:CRE1}
\end{equation}
where the collisionality was evaluated at mid-radius ($r/a \approx 0.5$), and the $q$ profile used for this analysis is $q = 2.1 + 2(r/a)^2$. Due to the strong dependence of the pitch-angle scattering rate $\nu_D$ on $v_{crit}$ we will need to employ an accurate description of how $v_{crit}$ varies with the system parameters. While the energy at which collisional drag is balanced by electric field acceleration provides an order of magnitude estimate of the critical speed for electron acceleration, a more refined estimate can be made by identifying the X-point energy \cite{guo2017phase}. Specifically, defining the energy and pitch fluxes by
\begin{equation}
\Gamma_p = \left[-\xi\frac{E_\Vert}{E_c} - C_F - \alpha p\gamma(1 - \xi^2)\right]f_e
, \label{gammaP}
\end{equation}
\begin{equation}
\Gamma_{\xi} = -\sqrt{1-\xi^2}\left(\frac{E_\Vert}{E_c} - \alpha\frac{p}{\gamma}\xi + \frac{p}{2}\nu_D\frac{\partial\ln f_e}{\partial\xi}\right)f_e
, \label{gammaXi}
\end{equation}
where $C_F$ is the collisional drag, $\gamma$ is the Lorentz factor, $\alpha \equiv \tau_c/\tau_s$, and $\tau_s = 6\pi\epsilon_0m_e^3c^3/(e^4B^2)$ defines the synchrotron radiation timescale. The form of the collisional drag and pitch-angle scattering coefficients employed in this work, including partial screening effects, is given in Ref. \cite{hesslow:2017}. For cases when the electric field is above the avalanche threshold, two nulls of the momentum space flux (i.e. $\Gamma_p=\Gamma_\xi = 0$) can be identified. The first is present at high energy and is distinct from, but correlated with, the bump in the steady state primary distribution~\cite{Decker:2016, guo2017phase}. This null in the momentum space flux, often referred to as an O-point, represents the energy at which the combination of drag, synchrotron radiation, and pitch-angle scattering balance electric field acceleration and thus provides the characteristic energy of the saturated primary distribution. The second null, referred to as the X-point, occurs at much lower energy for $E_\Vert \gg E_{av}$, and represents the separatrix between runaway and non-runaway orbits. Since the evaluation of the location of the X-point incorporates pitch-angle scattering and synchrotron radiation in addition to collisional drag, it provides a more accurate estimate of the threshold energy for electron acceleration compared to simply balancing collisional drag against electric field acceleration. We will thus use the X-point energy to estimate $v_{crit}$ in Eq. (\ref{eq:CRE1}), where the analytic model derived in Ref. \cite{mcdevitt2018relation} will be employed to efficiently estimate the location of the X-point energy. Section \ref{RPFSec} will utilize direct drift-kinetic simulations as a means of verifying this approximate analytic model.

\begin{figure}
\centering
\begin{subfigure}{.5\textwidth}
\caption{}
\includegraphics[width=\linewidth]{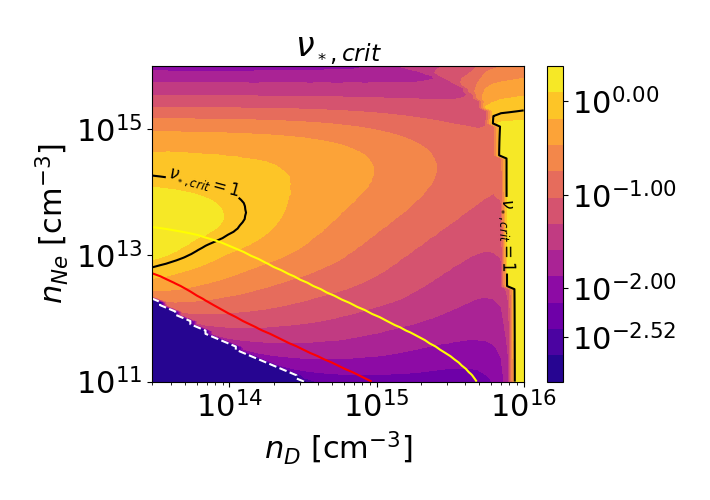}
\end{subfigure}%
\begin{subfigure}{.5\textwidth}
\caption{}
\includegraphics[width=\linewidth]{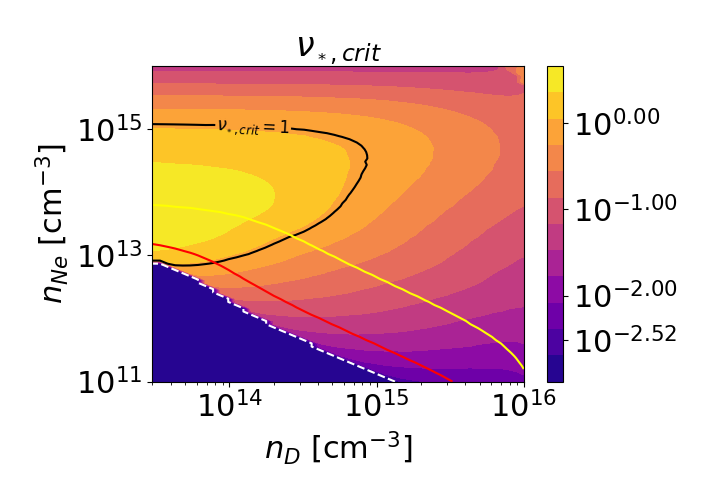}
\end{subfigure}
\begin{subfigure}{.5\textwidth}
\caption{}
\includegraphics[width=\linewidth]{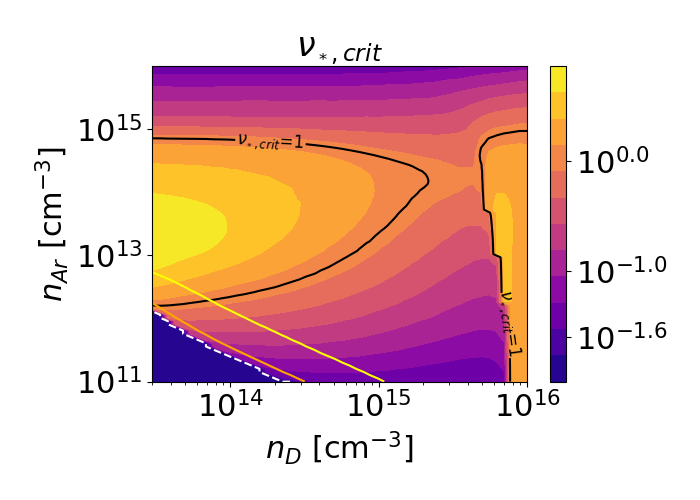}
\end{subfigure}%
\begin{subfigure}{.5\textwidth}
\caption{}
\includegraphics[width=\linewidth]{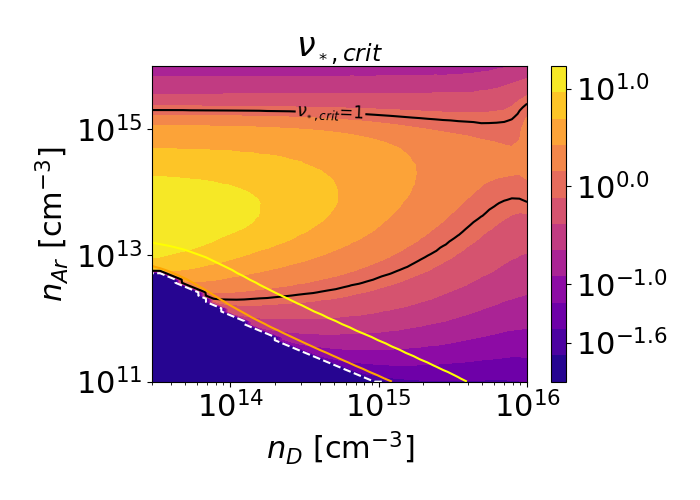}
\end{subfigure}
\caption{Collisionality evaluated at the critical energy ($\nu^*_{crit}$) for neon [panels (a) and (b)] and argon [panels (c) and (d)], where $j_{\parallel} = 1$ MA/m$^2$ for panels (a) and (c) and $j_{\parallel} = 2$ MA/m$^2$ for panels (b) and (d). The yellow, orange, and red contours represent 10,17, and 20 eV respectively. The black contour represents $\nu^*_{crit} = 1$, and the white dashed contour represents the region where the system is below the avalanche threshold ($E_\Vert/E_{av} < 1$). Other parameters used were $B = 5.3$ T, $q = 2.1 + 2(r/a)^2$, $R_0  = 6$ m, and $a = 2$ m.}
\label{nustar}
\end{figure}

The collisionality at the critical energy to run away $\nu_{*, crit}$ for plasmas with different compositions and $j_\Vert$ is shown in Figure \ref{nustar}. We see that $\nu_{*, crit}$ peaks similarly to $E_\Vert/E_c$. As the plasma composition transitions from $n_{impurity}/n_D \approx 1$ to $n_D \gg n_{impurity}$ in the region where 10 eV < $T_e$ < 20 eV (the red, yellow, and orange contours), $\nu_{*, crit}$ spans the entire collisionality regime.  In the regime where $n_{impurity}/n_D \approx 1$ the collisional regime ($\nu_{*, crit} \gtrsim 1$) is present, and we can anticipate that impurities robustly produce a collisional regime near the critical energy, thus reducing toroidal trapping effects. As the plasma composition approaches $n_D \gg n_{impurity}$, $\nu_{*, crit}$ approaches $\nu_{*, crit} \ll 1$, suggesting that initially trapped electrons will complete many bounce orbits before collisionally de-trapping. We also see in Figs. \ref{nustar}(a) and (c) that a second region where $\nu_{*, crit} \gtrsim 1$ exists, specifically at $n_D \approx 10^{16}$cm$^{-3}$, similarly to Figs. \ref{EoverEc} (a) and (c). The critical deuterium level for plasma recombination is reached in this regime, thus enhancing $E_\Vert/E_c$ and $\nu_{*, crit}$. At larger $j_\Vert$ (Figs. \ref{nustar} (b) and (d)), the critical deuterium density for recombination is increased, and the recombination region is also shifted to higher $n_D$. Larger $j_\Vert$ increases $\nu_{*, crit}$ across the entire range of plasma compositions, which arises from $E_\Vert \propto j_\Vert$ at a given plasma composition.
\section{Drift-Kinetic Solver for RE simulation}
\label{RAMc}
This section provides a brief description of the drift-kinetic code RAMc that will be used to directly evaluate the avalanche growth rate at different radial locations and for different plasma compositions [further details of the code can be found in Ref.  \cite{mcdevitt2019avalanche}]. The governing electron description used is
\begin{equation}
\frac{\partial f}{\partial t}  + \frac{d\mathbf{x}}{dt}\cdot\frac{\partial f}{\partial \mathbf{x}} + \frac{dp}{dt}\frac{\partial f}{\partial p} + \frac{d\xi}{dt} \frac{\partial f}{\partial \xi} = C(f) + C_{rad}(f)
\end{equation}
where time is normalized to $\tau_c$ and the relativistic guiding center from Ref. \cite{brizard2004guiding} are used. Synchrotron radiation's effect on runaway physics\cite{Decker:2016, guo2017phase} is accounted for in $C_{rad}(f)$. The collisions are modeled in $C(f)$, where $C(f) = C_{SA} + C_{LA}$. Here $C_{SA}$ is the small-angle collision operator and $C_{LA}$ is the large-angle collision operator. The small-angle collisions are described by the monte-carlo equivalents of the Lorentz and energy scattering operators~\cite{mcdevitt2019avalanche}. Large-angle collisions are captured through a M\"oller source term \cite{Moller:1932}, where both free and bound electrons are considered in the bulk electron density when generating secondary runaway electrons.
\begin{table}
\scriptsize
\begin{subtable}{0.45\textwidth}
\caption*{(a)}
\centering
\begin{tabular}{|c|c|c|c|c|} 
\hline
 $n_D$ [cm$^{-3}]$     &3$\times$10$^{13}$     &   6.4$\times$10$^{13}$   &   2.9$\times$10$^{14}$   &   2$\times$10$^{15}$  \\ \hline
 $n_{Ne}$ [cm$^{-3}]$  &4$\times$10$^{13}$  &   8.4$\times$10$^{12}$   &   1.5$\times$10$^{12}$ &   2$\times$10$^{11}$  \\ \hline
 $T_e$                 &10.48                  & 16.77                  & 16.2                  & 15.65                 \\ \hline
$\overline{Z}_{Ne}$    &   2.9919              &  3.977                 &3.912                 &3.909                  \\ \hline
$E/E_c$                &    783.9              &  464.1                 &84.17                   &12.4                   \\ \hline
$E/E_{av}$             &129.5                  &122.3                   &37.22                      & 6.378                 \\ \hline
$\mathbf{\nu^*_{crit}}$&      \textbf{3.236}   &\textbf{1.055}          &\textbf{0.09868}        &\textbf{0.01052}       \\ \hline
\end{tabular}
\end{subtable}
\hfill
\begin{subtable}{0.45\textwidth}
\centering
\caption*{(b)}
\begin{tabular}{|c|c|c|c|c|} 
\hline
$n_D$ [cm$^{-3}]$     &3$\times$10$^{13}$      &1.15$\times$10$^{14}$   &6.5$\times$10$^{14}$    &2.9$\times$10$^{15}$\\ \hline
$n_{Ar}$ [cm$^{-3}]$  &8$\times$10$^{12}$   &2$\times$10$^{12}$    &3$\times$10$^{11}$  &3.2$\times$10$^{10}$\\ \hline
$T_e$                 &11.39                   & 11.08                  & 11                & 13.82           \\ \hline
$\overline{Z}_{Ar}$   &   3.985               &  3.942                & 4.0777                &4.9206 \\ \hline
$E/E_c$               &    1,639               &  402.9                 &64.96                  &10.3 \\ \hline
$E/E_{av}$            &213.1                   &127                   &30.49                     & 5.29\\ \hline
$\mathbf{\nu^*_{crit}}$&      \textbf{7.04}   &\textbf{0.9918}          &\textbf{0.1136}       &\textbf{0.009607} \\ \hline
\end{tabular}
\end{subtable}
\caption{Summary of parameters used for the drift-kinetic simulations in Figs. \ref{RPFs}, \ref{avalanchee}, and \ref{psi10s} for a range of neon (a) and argon (b) densities, with $j_{\parallel} =$ 1.5 MA/m$^2$.}
\label{summaryValsjtot2}
\end{table}

\section{Results}
\label{results}
\subsection{Runaway Probability Function}
\label{RPFSec}
The runaway probability function (RPF) provides a convenient means of assessing how tokamak geometry modifies the efficiency of the avalanche mechanism across a range of plasma compositions. Specifically, the RPF~\cite{karney1986current, Liu:2016, zhang2017backward, mcdevitt2023physics} indicates the probability that an electron with a given initial phase space location will run away at a later time. By considering how this function varies across different plasma compositions this will provide direct insight into which regions of phase space contribute to the avalanche of runaway electrons. To evaluate the RPF, we will utilize the drift-kinetic solver described in Sec. \ref{RAMc}, but with the large-angle collision operator turned off, and a large number of marker particles used to sample the region of interest. Over time marker particles will either accelerate to relativistic energies, or slow down to the bulk plasma. By keeping track of the marker particle's initial location, and its final energy, the probability of electrons running away can then be computed. 

\begin{figure}
\centering
\begin{subfigure}{.5\textwidth}
\caption{}
\includegraphics[width=\linewidth]{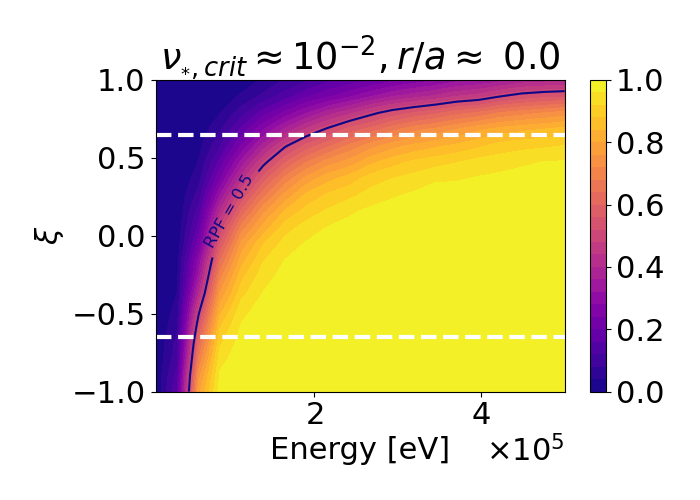}
\end{subfigure}%
\begin{subfigure}{.5\textwidth}
\caption{}
\includegraphics[width=\linewidth]{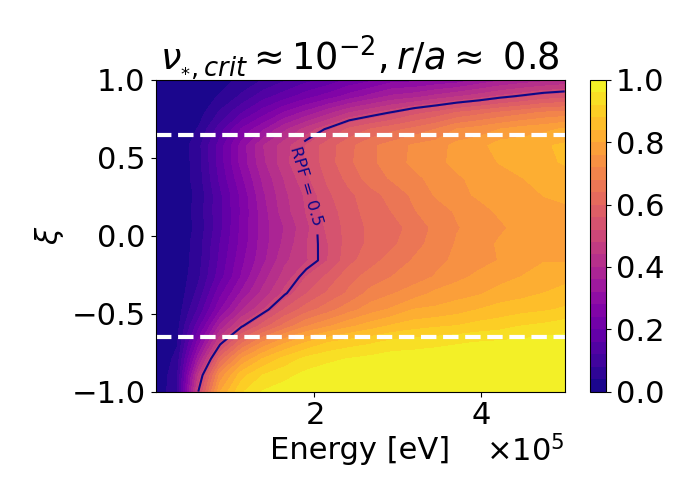}
\end{subfigure}
\begin{subfigure}{.5\textwidth}
\caption{}
\includegraphics[width=\linewidth]{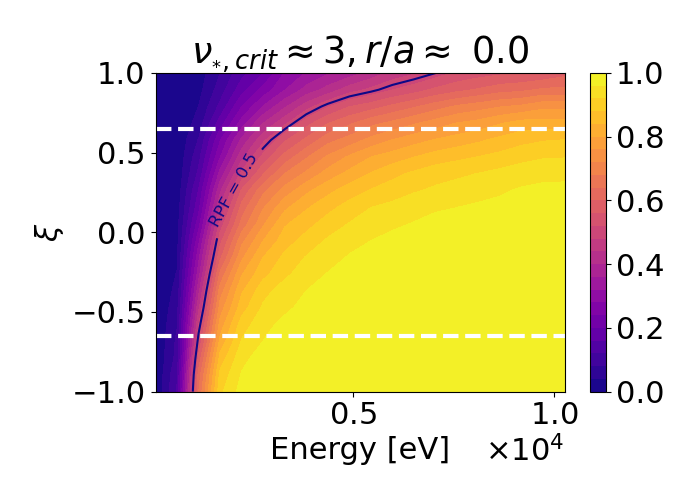}
\end{subfigure}%
\begin{subfigure}{.5\textwidth}
\caption{}
\includegraphics[width=\linewidth]{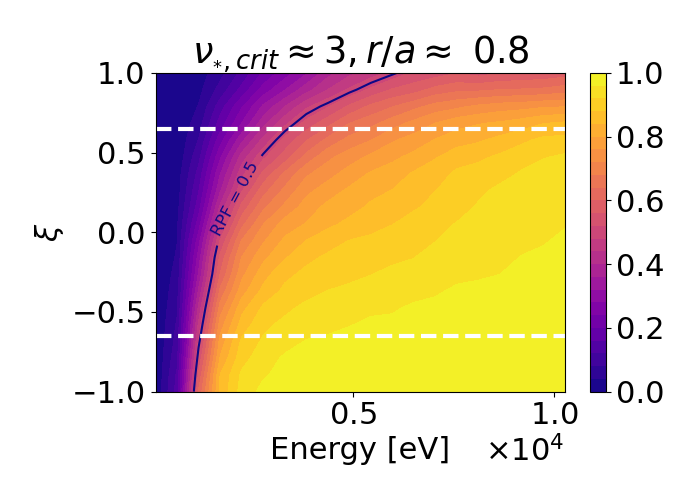}
\end{subfigure}
\caption{Runaway probability distribution for a collisionless [panels (a) and (b)] and collisional [panels (c) and (d)] post TQ plasma with $n_D$ = 2 $\times$ 10$^{15}$ cm$^{-3}$ and $n_{Ne}$ = 2 $\times$ 10$^{11}$ cm$^{-3}$  for figures (a) and (b), and $n_D$ = 2 $\times$ 10$^{13}$ cm$^{-3}$, $n_{Ne}$ = 3 $\times$ 10$^{13}$ cm$^{-3}$ for panels (a) and (b). Panels (a) and (c) are on axis and panels (b) and (d) are on the outboard side. The black contour represents the separatrix, where the critical energy and $\xi$ for REs are, and the white contours represent the trapped-passing boundary. Input parameters were $B = 5.3$ T, $j_{\parallel} = $1.5 MA/m$^2$, $q = 2.1 + 2(r/a)^2$, $R_0  = 6$ m, and $a = 2$ m.}
\label{RPFs}
\end{figure}

The RPFs for two different plasma compositions and two different radii are shown in Fig. \ref{RPFs}. One million electrons are initialized with energies near the critical energy obtained from the O-X merger model and $\xi \in (-1,1)$. Here, the black contours are the 50\% contour and the white contours indicate the trapped-passing boundary. We have selected a plasma composition with a substantial quantity of impurities and thus a high $\nu_{*,crit} \gtrsim 1$ [Figs. \ref{RPFs}(c) and (d)] and a deuterium-dominated plasma with $\nu_{*,crit} \ll 1$ [Figs. \ref{RPFs}(a) and (b)].

One immediate result of the RPF calculation is the verification of the critical energy estimation from the O-X merger model, where the critical energy can be taken to be the RPF = 0.5 contour at $\xi = -1$ in Figure \ref{RPFs}. 
Indeed, it was found that the RPF = 0.5 contour at $\xi = -1$ was within 14\% of the critical energy provided by the O-X merger model. Turning to Figs. \ref{RPFs} (a) and (b), these RPFs are in the low collisionality regime, where an expected reduction in the RPF is present for trapped electrons as $r/a$ increases (compare the $r/a\approx 0$ and $r/a\approx 0.8$ cases). However, in the collisional regime, those electrons that would typically be trapped at $\nu_{*, crit} \ll 1$ are now not trapped, as shown in Figure \ref{RPFs} (d). In fact, the RPF is minimally impacted at $\nu_{*, crit} \gtrsim 1$, highlighting that electrons in the trapped region indeed collisionally de-trap and become REs. These collisionally de-trapped electrons can become runaways, which is anticipated to enhance the avalanche efficiency. Moreover, in the $\nu_{*, crit} \gtrsim 1$ regime, the energy where RPF = 0.5 and $\xi = -1$ decreases by over an order of magnitude, allowing more electrons to enter the runaway regime and generate secondary electrons via avalanche. 

\begin{figure}
\centering
\begin{subfigure}{.5\textwidth}
\caption{}
\includegraphics[width=\linewidth]{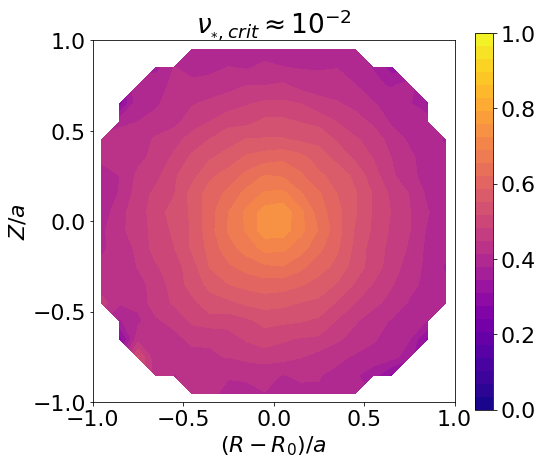}
\end{subfigure}%
\begin{subfigure}{.5\textwidth}
\caption{}
\includegraphics[width=\linewidth]{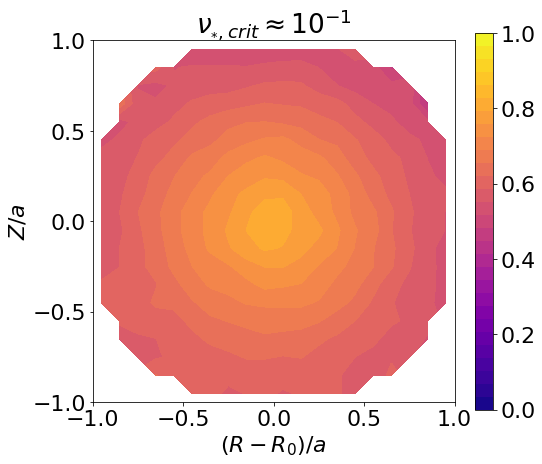}
\end{subfigure}
\begin{subfigure}{.5\textwidth}
\caption{}
\includegraphics[width=\linewidth]{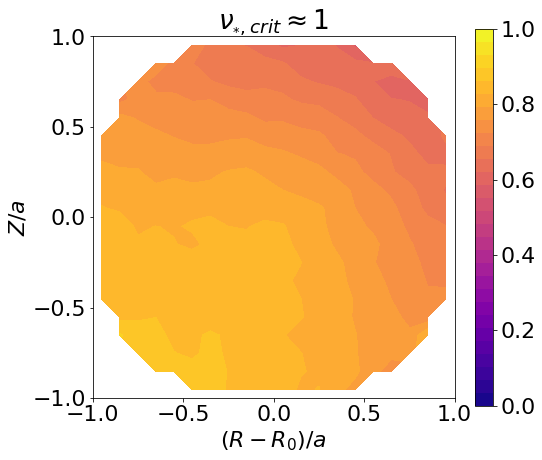}
\end{subfigure}%
\begin{subfigure}{.5\textwidth}
\caption{}
\includegraphics[width=\linewidth]{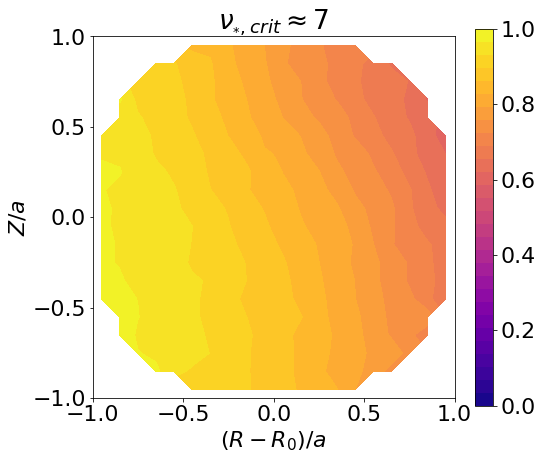}
\end{subfigure}
\caption{RPF projected on to the R-Z plane for argon at varying $\nu_{*, crit}$. 10$^6$ maker particles were used and initialized with energy $K = K_{crit}$ and $\mu = 0$ ($\xi \approx$ -0.9 to -1.0 on the weak field side. Simulation parameters can be found in Table \ref{summaryValsjtot2}, and other parameters used are $B = 5.3$ T, $j_{\parallel} = $1.5 MA/m$^2$, $q = 2.1 + 2(r/a)^2$, $R_0  = 6$ m, and $a = 2$ m.}
\label{RPFspatial}
\end{figure}

The impact of $\nu_{*, crit}$ can also be seen in spatial coordinates, where $R$ and $Z$ will be the radial and axial coordinates. In the collisionless limit, electrons can be described by their invariants $f_e \approx f_e(K, \mu, p_\phi)$, which are the electrons' energy $K$, magnetic moment $\mu$, and toroidal canonical momentum $p_\phi$. For small $\rho^*_e \equiv \rho_e/a$, where $\rho_e$ is the electron gyroradius, $p_\phi$ becomes dominated by the poloidal flux function $\psi$. Therefore, the electron distribution function can be approximated as $f_e \approx f_e(\gamma, \mu, \psi)$.  The RPF at different $\nu_{*, crit}$ is shown in Figure \ref{RPFspatial} for argon impurities, where electrons are initialized near the critical energy and with $\mu = 0$ ($\xi \approx$ -0.9 to -1.0 on the weak field side). In Figure \ref{RPFspatial} we see in panel (a) ($\nu_{*, crit} \approx 10^{-2}$) that the RPF contours indeed follow $\psi$, however, moving from panel (a) to panel (d), where $\nu_{*, crit}$ increases from $10^{-2}$ to 7, the RPF structure changes significantly. Between panels (a) and (b), electrons continue to largely follow the flux surfaces, but as $\nu_{*, crit}$ increases from 10$^{-1}$ to 1 [panels (b) and (c)], the RPF structure departs from following flux surfaces. The modified RPF structure results from electrons on the outer flux surfaces collisionally de-trapping and becoming passing electrons. For larger $\nu_{*, crit}$ [panel (d)], electrons have the largest probability of running away on the inboard side, and the RPF structure now follows roughly a $1/R$ function, which results from the form of the electric field $E_\Vert = R_0E_1/R$, where $R_0$ is the major radius and $E_1$ is the electric field at the magnetic axis ($R = R_0$). The modified RPF structure thus follows $E_\Vert$ in the limit of $\nu_{*, crit} > 1$, where electrons across the $R$-$Z$ plane are largely passing. Since $E_\Vert$ is strongest on the inboard side, electrons thus have the largest probability of running away there, and the RPF is the largest.

\subsection{Avalanche Growth Rate Evaluation}
\label{avalanche}

\begin{figure}
\centering
\begin{subfigure}{.45\textwidth}
\caption{}
\includegraphics[width=\linewidth]{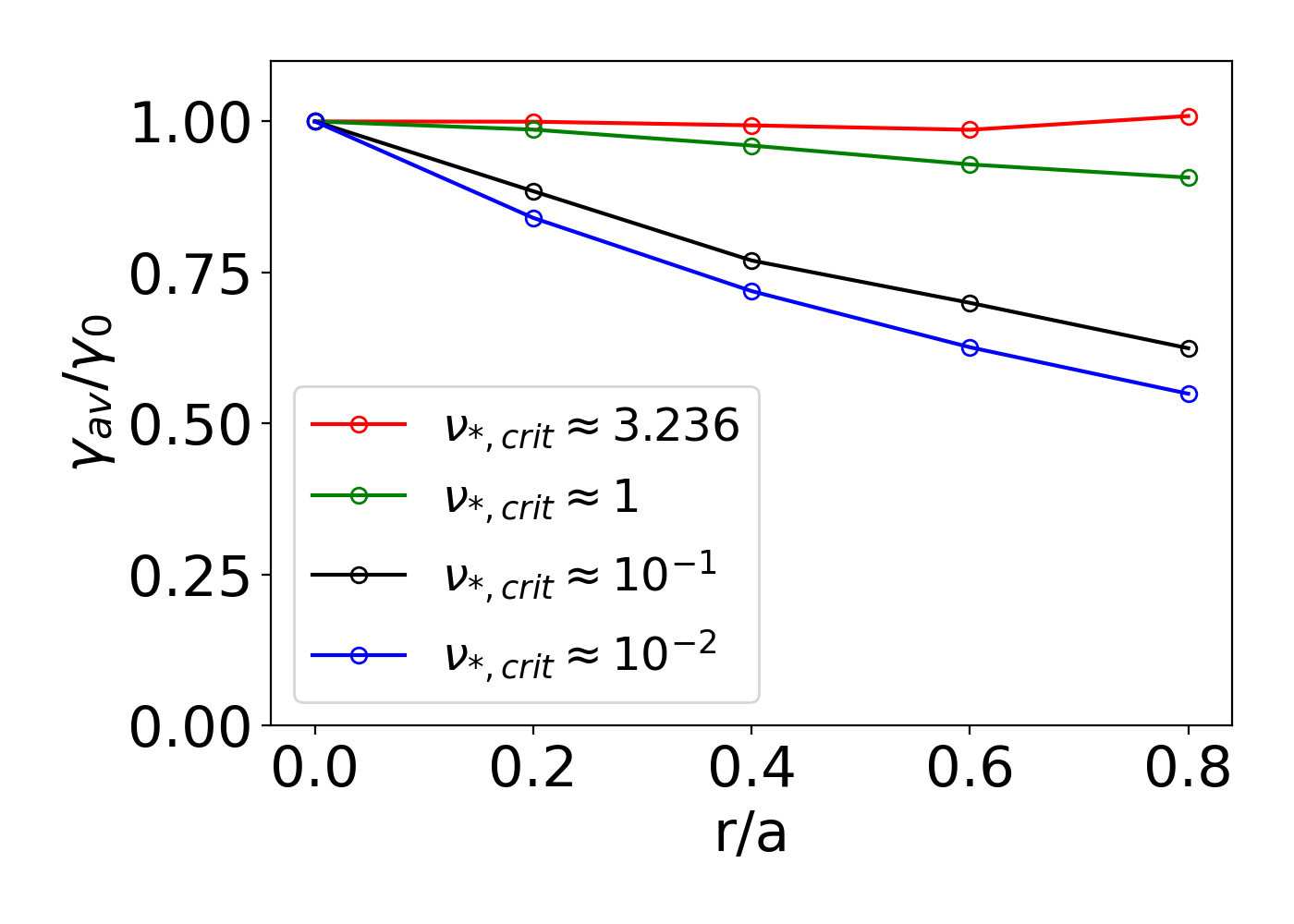}
\end{subfigure}
\begin{subfigure}{.45\textwidth}
\caption{}
\includegraphics[width=\linewidth]{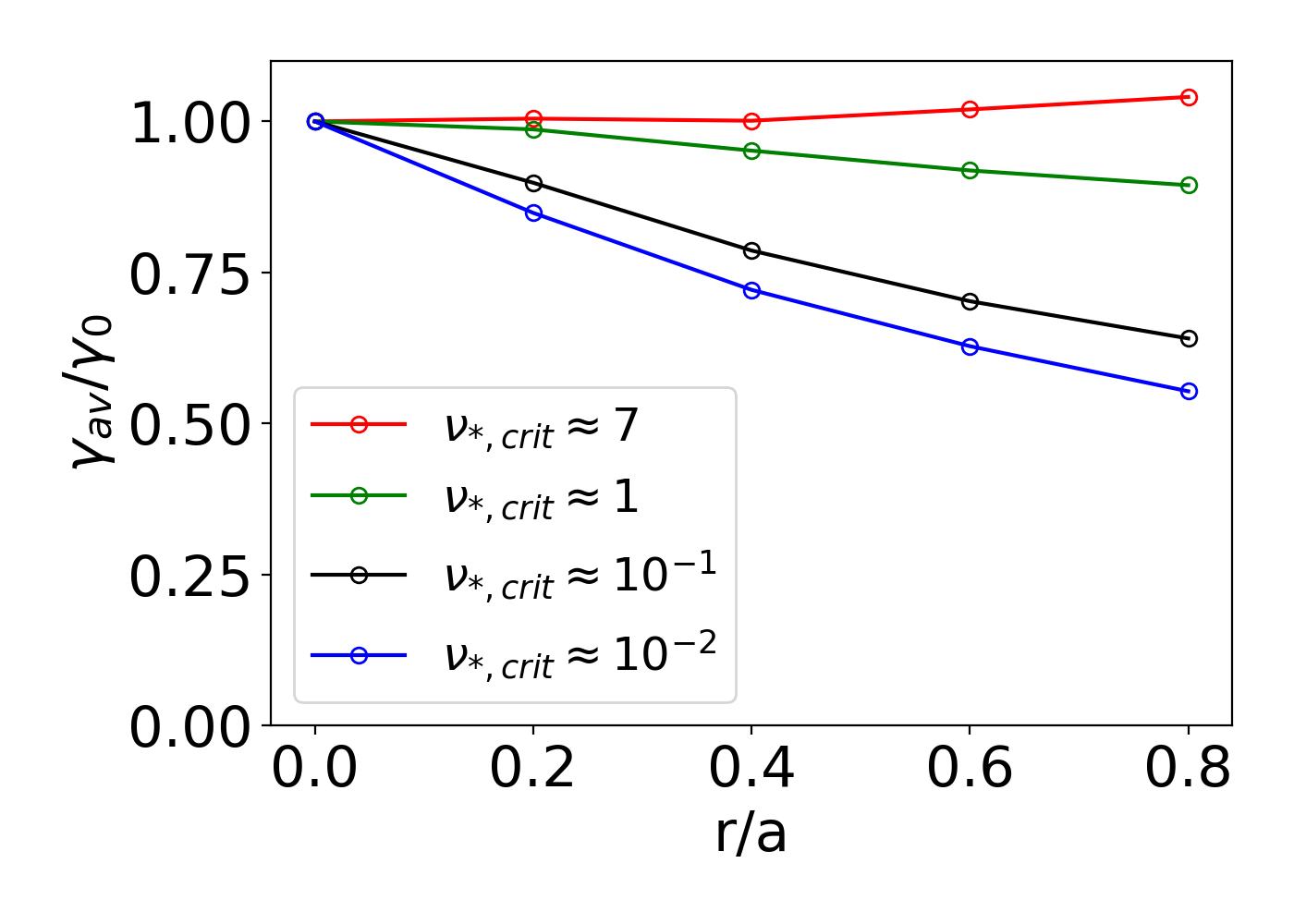}
\end{subfigure}
\caption{Avalanche growth rates, normalized to the on axis value for neon (a) and argon (b). The dashed lines represent cases where synchrotron radiation was turned off, and $B$ = 53 T. The solid lines represent cases where synchrotron radiation was kept on, and $B$ = 5.3T. The full list of input parameters can be found in Table \ref{summaryValsjtot2}.}
\label{avalanchee}
\end{figure}

Quantifying the impact of $\nu_{*,crit}$ on the avalanche mechanism is done by directly evaluating the avalanche growth rate $\gamma_{av}$ as a function of $r/a$ using the drift kinetic solver described in Sec. \ref{RAMc}. Here, a small initial seed population of REs is used to initiate the exponential growth. The avalanche growth rate is evaluated as a function of $r/a$ at varying $\nu_{*, crit}$ for both argon and neon and is shown in Fig. \ref{avalanchee}, with the tabulated parameters in Table \ref{summaryValsjtot2}.  

In the collisionless ($\nu_{*, crit} \approx 10^{-2}$) region, the avalanche growth rate is reduced by almost 50\% at $r/a = 0.8$. However, as $\nu_{*, crit}$ increases to the collisional regime, the reduction of the avalanche growth rate at large $r/a$ becomes smaller, up to the point where it can be considered negligible, contrary to widely used formulations of the avalanche growth rate\cite{Rosenbluth:1997,Chiu:1998}.  The largest impact from impurities on the avalanche growth rate is between $\nu_{*, crit} \approx 0.1 - 1$, where the radial reduction varies anywhere from $\sim 40\%$ to $\sim 10\%$. 

\begin{figure}
\centering
\includegraphics[scale=0.5]{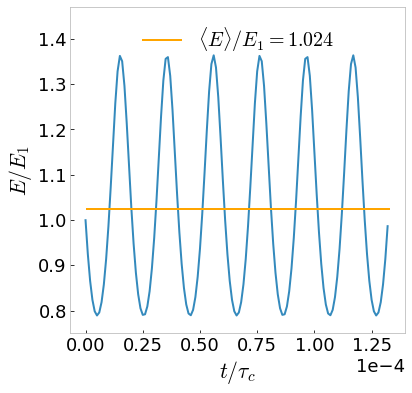}
\caption{Simulation of the toroidal electric field seen by an electron as a function of time, where time is normalized to $\tau_c$. Simulation parameters used here were $B = 5.3$ T, $q = 2.1 + 2(r/a)^2$, $R_0  = 6$ m, and $a = 2$ m. The blue curve represents the electric field the electron sees at a given poloidal location normalized to $E_1$, and the orange curve represents the average of this quantity over time.}
\label{EpolTransit}
\end{figure}

One feature present is in the case of argon, where $\nu_{*,crit} \approx 7$. It is seen that the avalanche growth rate increases slightly with radius. The increased avalanching on the outboard side has been noted in the analysis of Dreicer RE generation in Ref.\cite{mcdevitt2019runaway}, however here it is seen in the context of avalanche generation, albeit with only a very modest increase evident. This slight increase is due to the average electric field seen by electrons are large radii slightly exceeding the on-axis value. In particular, if the electric field $E_\Vert$ is expanded at the inboard ($R - r$) and outboard ($R + r$) sides, $E(R_0 -  r) \approx E_{1}(r)[1 + (r/R_0) +  (r/R_0)^2 + \dots]$ for the inboard side and $E(R_0 + r) \approx E_{1}(r)[1 - r/R_0 +  (r/R_0)^2 + \dots]$ for the outboard side. While the first order terms cancel, implying that the electric field an electron sees averages out as it does a poloidal transit, the second order terms do not. As a result electrons at large $r/a$ can see an average electric field enhanced by a factor of $(r/R_0)^2$. This feature can be seen by evaluating the average electric field an electron sees as it does many poloidal transits. Figure \ref{EpolTransit} illustrates the instantaneous and average electric fields an electron located at $r/a = 0.8$ sees as it completes several poloidal transits. Here, the oscillations in the instantaneous electric field seen by the electron are due to the $1/R$ dependence of the electric field; however, as indicated by the orange line, the average value of the electric field is slightly greater than one ($\left\langle E \right\rangle / E_1 \approx 1.024$). While this enhancement of the electric field is modest, the RE avalanching deviates from a linear scaling with electric field strength for $E_\Vert/E_c \gg 1$, resulting in enhanced avalanching at large $r/a$, where the enhancement at $r/a \approx 0.8$ was $\sim$ 4\%.

\subsection{Efficiency of the Avalanche Mechanism}

A powerful means of estimating the maximum number of RE exponentiations possible for a given disruption scenario is by predicting the amount of poloidal flux that must be consumed to effect an order of magnitude amplification of a given seed RE population. Such an estimate directly measures the efficiency through which the avalanche process converts poloidal flux into increases in the RE population, often a more useful metric than specific values of the avalanche growth rate at given parameters. In particular, taking the avalanche growth rate to scale linearly with the electric field strength~\cite{Rosenbluth:1997}, and assuming the electric field is much greater than the avalanche threshold, allows the avalanche growth rate to be expressed as:
\begin{equation}
\gamma_{av} = \gamma_{exp} \left( \frac{E_\Vert}{E_c} - 1\right) \approx \frac{\gamma_0 E_\Vert}{E_c}
, \label{eq:EAM1}
\end{equation}
where $\gamma_{exp}$ is a constant coefficient that determines the efficiency of the avalanche. Noting that the inductive electric field is related to the change in the poloidal flux via the relation $E_\Vert \approx \left( 1/R_0\right) \partial \psi/\partial t$, where $\psi$ is the poloidal flux function, Eq. (\ref{eq:EAM1}) can be rewritten as
\begin{equation}
\gamma_{av} \approx \frac{\gamma_{exp}}{E_cR_0 } \frac{\partial \psi}{\partial t}
. \label{eq:EAM2}
\end{equation}
Integrating this expression over the time interval $t_i$ to $t_f$, yields the total number of exponentiations of the RE population, i.e.
\begin{equation}
N_{\text{exp}} = \int^{t_f}_{t_i} dt \gamma_{av} \approx \frac{\gamma_{exp}}{E_cR_0 }  \int^{t_f}_{t_i} dt \frac{\partial \psi}{\partial t} =  \frac{\gamma_{exp}}{E_cR_0 } \left[ \psi \left( t_f\right) - \psi \left( t_i\right) \right] = \frac{\gamma_{exp}}{E_cR_0} \Delta \psi
, \label{eq:EAM3}
\end{equation}
where $\Delta \psi \equiv \psi \left( t_f\right) - \psi \left( t_i\right)$ and we have assumed the plasma composition to be unchanged such that $\gamma_{exp}/\left( E_c R_0\right)$ is constant. The number of exponentiations of the RE population is thus directly linked to the change of the poloidal flux, the coefficient $\gamma_{exp}$, and the parameters $E_c$ and $R_0$. In addition, it is evident that the number of expected avalanche exponentiations does not depend on the specifics of the disruption trajectory, only the change of the poloidal flux so long as the coefficient $\gamma_{exp}/E_cR_0$ remains fixed. If we take the familiar expression for $\gamma_{exp}$ derived in Ref. \cite{Rosenbluth:1997}, 
\[
\gamma_{exp} \approx \frac{1}{\tau_c \ln \Lambda} \sqrt{\frac{\pi \varphi}{3 \left( Z_{eff}+5\right)}}
,
\]
where $\varphi \approx \left( 1+1.46\sqrt{\varepsilon} + 1.72 \varepsilon \right)$, this allows Eq. (\ref{eq:EAM3}) to be written as
\begin{equation}
N_{\text{exp}} \approx \frac{1}{R_0 E_c \tau_c \ln \Lambda} \sqrt{\frac{\pi \varphi}{3 \left( Z_{eff}+5\right)}} \Delta \psi = \frac{\Delta \psi}{\psi_{\text{exp}}}
. \label{eq:EAM4}
\end{equation}
Here, we have introduced the quantity $\psi_{\text{exp}}$ defined by
\begin{equation}
\frac{1}{\psi_{\text{exp}}} \equiv  \frac{1}{R_0 E_c \tau_c \ln \Lambda} \sqrt{\frac{\pi \varphi}{3 \left( Z_{eff}+5\right)}} = \frac{e}{m_e c} \sqrt{\frac{\pi \varphi}{3 \left( Z_{eff}+5\right)}} \frac{1}{R_0} \frac{1}{\ln \Lambda}
, \label{eq:EAM5}
\end{equation}
which is a constant that determines the number of exponentiations a RE population undergoes for a given change of poloidal flux. Equation (\ref{eq:EAM5}) implies the efficiency of the RE avalanche is only sensitive to a modest number of quantities, primarily the Coulomb logarithm $\ln \Lambda$, the effective charge $Z_{eff}$ and the radial location through $\varepsilon = r/R_0$. The major radius dependence indicated in Eq. (\ref{eq:EAM5}) will cancel, due to the poloidal flux scaling with the major radius. For example, in circular geometry the poloidal flux function can be expressed in terms of the plasma current by 
\begin{equation}
\psi(r) = \frac{-R_0\mu_0}{2\pi}\int_r^a \frac{I_p(r^{\prime})}{r^{\prime}}dr^{\prime}
, \label{eq:EAM6}
\end{equation}
where $I_p \left( r \right)$ indicates the plasma current inside a radius $r$. A convenient normalization of $\psi_{\text{exp}}$ will thus be $\bar{\psi}_{\text{exp}} \equiv 2\pi \psi_{\text{exp}} / \left( \mu_0 R_0 \right)$, which removes the major radius scaling, and has units of current. The quantity $\bar{\psi}_{\text{exp}}$ thus provides a rough measure of the drop of the plasma current required to effect one exponentiation of the RE population. A closely related, but more convenient quantity, is $\bar{\psi}_{10} = \ln 10 \bar{\psi}_{\text{exp}}$, which is directly related to the drop in plasma current required to induce an order of magnitude increase of the RE population  via avalanching~\cite{Boozer:2018}. Defining the Alfven current by $I_A \equiv 4\pi m_e c / \left( \mu_0 e\right) \approx 0.017\;\text{MeV}$, $\bar{\psi}_{10}$ can be expressed as:
\begin{equation}
\bar{\psi}_{10} = \frac{\ln 10}{2} I_A \sqrt{\frac{3 \left( Z_{eff}+5\right)}{\pi \varphi}} \ln \Lambda
. \label{eq:EAM7}
\end{equation}
For a pure deuterium plasma, taking the RE beam to be tighly localized about the magnetic axis such that $\varphi \approx 1$, and noting that the Coulomb logarithm seen by relativistic electrons will exceed the Coulomb logarithm for thermal electrons~\cite{Solodov:2008}, such that we may take $\ln \Lambda = 20$, Eq. (\ref{eq:EAM7}) implies $\bar{\psi}_{10} \approx 0.94\; \text{MA}$, indicating that a tokamak with $15\;\text{MA}$ of current, would have an avalanche gain of roughly $\sim 10^{15/0.94} \approx 10^{16}$, which is the often cited expected avalanche gain of ITER~\cite{Hender:2007}. 

\begin{figure}
\centering
\begin{subfigure}{.45\textwidth}
\caption{}
\includegraphics[width=\linewidth]{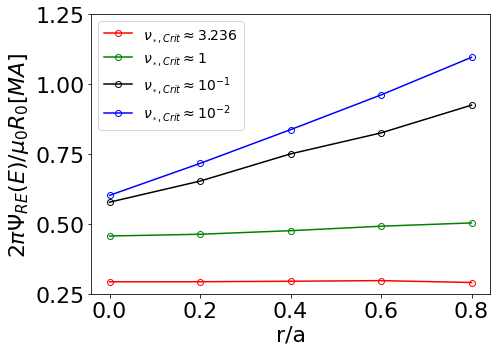}
\end{subfigure}
\begin{subfigure}{.45\textwidth}
\caption{}
\includegraphics[width=\linewidth]{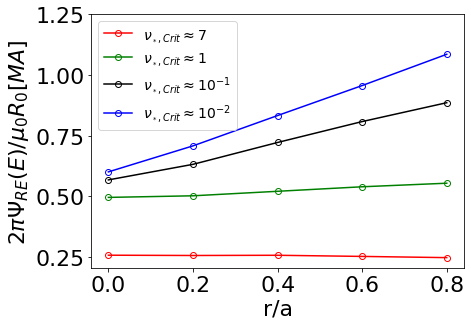}
\end{subfigure}
\caption{Efficiency of the avalanche mechanism as a function of radius at different collisionalities for neon (a) and argon (b). The parameters are in Table \ref{summaryValsjtot2}.}
\label{psi10s}
\end{figure}

While Eq. (\ref{eq:EAM7}) provides a useful baseline for estimating the avalanche potential for a range of tokamak concepts, several uncertainties in the appropriate value of $\bar{\psi}_{10}$ remain. In particular, the toroidal factor $\varphi$ implies that the avalanche growth rate for a given $E_\Vert$ can be reduced by roughly a factor of two at large minor radii. While it may often be the case that RE beams are largely localized to the inner half of the plasma, even a thirty percent decrease in $\gamma_{exp}$ due to toroidal corrections, would reduce the avalanche gain in ITER to $10^{15/1.22} \approx 2 \times 10^{12}$, substantially reducing the importance of nuclear seeding mechanisms~\cite{Martin:2017}. As indicated in Fig. \ref{avalanchee}, this reduction due to toroidicity is only expected for plasmas with $\nu_{*, crit} < 1$, which tend to be deuterium dominated plasmas. The expected $\bar{\psi}_{10}$ for $\nu_{*, crit} < 1$ plasmas is shown in Fig. \ref{psi10s} (blue and black curves) as a function of minor radius. Here it is evident that a substantial increase of $\bar{\psi}_{10}$ is present as $r$ is increased, though the on-axis value is substantially less than the often invoked value of $\bar{\psi}_{10} \approx 0.94\;\text{MA}$. This is due to Ref. \cite{Hender:2007} choosing a Coulomb logarithm of $\ln \Lambda = 20$, which is larger than the Coulomb logarithm estimated from the more recent expressions derived in Refs. \cite{Solodov:2008, hesslow:2017}, particularly for the high deuterium densities that are required to maintain a plasma temperature of roughly ten to twenty eV in a plasma with modest impurity content.

In addition to toroidal corrections, Ref. \cite{hesslow2019influence} showed that the presence of partially ionized impurities led to a strongly nonlinear scaling of the avalanche growth rate, leading to a substantial increase in the efficiency of the RE avalanche at large electric fields. The presence of this nonlinear scaling of the avalanche growth rate prevents a unique values of $\bar{\psi}_{10}$ from being evaluated since the time integration indicated in Eq. (\ref{eq:EAM3}) can no longer be analytically computed. Instead, the effective value of $\bar{\psi}_{10}$ will depend on the strength of the electric field, where larger electric fields will lead to a smaller value of $\bar{\psi}_{10} \left( E_\Vert \right)$. In Ref. \cite{mcdevitt2019avalanche} values near $\bar{\psi}_{10} \left( E_\Vert \right) \gtrsim 0.5\;\text{MA}$ were found on-axis for electric fields of tens of times the Connor-Hastie threshold. For the cases indicated in Table \ref{summaryValsjtot2}, $\gamma_{exp}$ can be evaluated via the expression:
\begin{equation}
\gamma_{exp} = \frac{\gamma_{av}}{(E_\Vert/E_c)-1}
,
\end{equation}
where $\gamma_{av}$ is the avalanche growth rate previously evaluated. The resulting values of $\bar{\psi}_{10} \left( E_\Vert \right)$ are shown in Fig. \ref{psi10s} for different minor radii. Perhaps the most striking feature in Fig. \ref{psi10s} lies in the region where $\nu_{*, crit} \gtrsim 1$ for both argon and neon, resulting in $\bar{\psi}_{10}$ values as low as $\bar{\psi}_{10} \approx 0.25$. The highly efficient avalanche at these parameters arises from the non-linear dependence of the avalanche growth rate on the electric field. From Table \ref{summaryValsjtot2}, the non-normalized electric field ranged from 10 V/m to 60 V/m (hundreds to over a thousand times the Connor-Hastie threshold), which is well within the non-linear region of the avalanche growth rate~\cite{hesslow2019influence}, resulting in the efficiency of the RE avalanche being substantially enhanced compared to Eq. (\ref{eq:EAM7}). An additional striking feature is that depending on the value of $\nu_{*, crit}$, $\bar{\psi}_{10}$ can vary from $\psi_{10} \approx 0.25 - 1.1$ at large minor radii, a far larger scatter in the efficiency of the RE avalanche than has been reported previously.

\section{Conclusions}
\label{discuss}

This work addresses the net impact of impurities and toroidal geometry on the RE avalanche mechanism for a range of parameters typical of tokamak disruptions. As RE avalanching is expected to be a large source of RE generation in future tokamaks such as ITER \cite{Hender:2007}, resolving uncertainties in this mechanism is imperative to identifying appropriate disruption mitigation strategies. The net impact from competing effects on the critical energy for RE generation produced by impurities is evaluated using a self-consistent power balance relation and an electric field estimated via Ohm's law. It was found that the strong radiative cooling resulting from substantial quantities of neon and argon impurities, and the subsequent large electric fields (see Fig. \ref{EoverEc}), leads to a decrease in the critical energy for an electron to run away. Consequently, electrons were found to be in the collisional regime ($\nu_{*, crit} \gtrsim 1$) for cases where a substantial inventory of impurities was present ($n_{impurity}/n_D \gtrsim 1)$. For these plasmas, the efficiency of the RE avalanche mechanism was found to be largely insensitive to magnetic trapping by the tokamak magnetic field. In contrast, as the plasma composition approaches a deuterium-dominant plasma, it is found that $\nu_{*, crit} \ll 1$. For this latter case the efficiency of the RE avalanche mechanism was found to drop as the radius was increased, consistent with previous results. The analysis in this work also shows that $\nu_{*, crit}$ serves as a guiding parameter to dictate the efficiency of the avalanche mechanism, in addition to indicating when toroidal trapping effects are prominent. 

\begin{acknowledgements}

This work was supported by the U.S. Department of Energy, Office of Science, Office of Fusion Energy Sciences under award No. DE-SC0024634. The authors acknowledge the University of Florida Research Computing for providing computational resources that have contributed to the research results reported in this publication.

\end{acknowledgements}

\newpage

\bibliographystyle{apsrev}
\bibliography{./ref.bib}

\end{document}